\documentclass[prd,twocolumn,tightenlines,showpacs,nofootinbib,superscriptaddress,nobalancelastpage]{revtex4-1}

\usepackage[usenames,dvipsnames,svgnames,table]{xcolor}

\usepackage{graphics}
\usepackage{graphicx}
\usepackage{longtable}
\usepackage{url}
\usepackage{bm}
\usepackage{natbib}
\usepackage{textpos}
\usepackage{amsfonts,amsmath,amssymb,mathrsfs,bm,amsthm}
\usepackage{float}
\restylefloat{table}
\usepackage{booktabs}
\usepackage{array}
\usepackage{tabu}
\usepackage{dcolumn}
\usepackage{rotating}
\usepackage{ulem}

\usepackage{amsmath}
\usepackage{amssymb}
\usepackage{graphicx}
\usepackage{epstopdf}
\usepackage{inputenc}

\usepackage{euscript}

\newcommand{\RN}[1]{
  \textup{\uppercase\expandafter{\romannumeral#1}}
}

\usepackage{nicefrac}
\usepackage{amsthm}
\usepackage{color}
\usepackage{cancel}
\usepackage{appendix}
\usepackage{makecell}

\usepackage[colorlinks=true,linkcolor=black,citecolor=blue,urlcolor=blue,bookmarksopen]{hyperref}

\newcommand{\appropto}{\mathrel{\vcenter{
  \offinterlineskip\halign{\hfil$##$\cr
    \propto\cr\noalign{\kern2pt}\sim\cr\noalign{\kern-2pt}}}}}
\renewcommand{\v}[1]{\boldsymbol{#1}}		

\def\prn#1{{\left(#1\right)}}

\def\sbrk#1{{\left[#1\right]}}

\def\bra#1{{\langle#1|}}

\def\cg(#1,#2)(#3,#4)(#5,#6){\bra{#1,#2,#3,#4}#5,#6\rangle}

\def\ts#1{{_{\mbox{\scriptsize #1}}}}

\def\threej(#1,#2)(#3,#4)(#5,#6){\begin{pmatrix}#1&#3&#5\\#2&#4&#6\end{pmatrix}}
\def\sixj(#1,#2,#3)(#4,#5,#6){\begin{Bmatrix}#1&#2&#3\\#4&#5&#6\end{Bmatrix}}
\def\ninej(#1,#2,#3)(#4,#5,#6)(#7,#8,#9){\begin{Bmatrix}#1&#2&#3\\#4&#5&#6\\#7&#8&#9\end{Bmatrix}}

\def\sV{{\ensuremath{\EuScript V}}}

\def\mb{\mathbf}

\begin{document}


\title{Ferromagnetic Gyroscopes for Tests of Fundamental Physics}


\date{\today}

\author{Pavel~Fadeev}
\email{pavelfadeev1@gmail.com}
\affiliation{Helmholtz-Institut, GSI Helmholtzzentrum f{\"u}r Schwerionenforschung, 55128 Mainz, Germany}
\affiliation{Johannes Gutenberg-Universit{\"a}t Mainz, 55128 Mainz, Germany}

\author{Chris~Timberlake}
\affiliation{School of Physics and Astronomy, University of Southampton, SO17 1BJ Southampton, United Kingdom}

\author{Tao~Wang}
\affiliation{Department of Physics, Princeton University, Princeton, New Jersey 08544, USA}

\author{Andrea~Vinante}
\affiliation{School of Physics and Astronomy, University of Southampton, SO17 1BJ Southampton, United Kingdom}
\affiliation{Istituto di Fotonica e Nanotecnologie – CNR and Fondazione Bruno Kessler, I-38123 Povo, Trento, Italy}

\author{Y.~B.~Band}
\affiliation{Department of Chemistry, Department of Physics, Department of Electro-Optics, and the Ilse Katz Center for Nano-Science, Ben-Gurion University, Beer-Sheva 84105, Israel}

\author{Dmitry~Budker}
\affiliation{Helmholtz-Institut, GSI Helmholtzzentrum f{\"u}r Schwerionenforschung, 55128 Mainz, Germany}
\affiliation{Johannes Gutenberg-Universit{\"a}t Mainz, 55128 Mainz, Germany}
\affiliation{Department of Physics, University of California at Berkeley, Berkeley, California 94720-7300, USA}

\author{Alexander~O.~Sushkov}
\affiliation{Department of Physics, Boston University, Boston, Massachusetts 02215, USA}

\author{Hendrik~Ulbricht}
\affiliation{School of Physics and Astronomy, University of Southampton, SO17 1BJ Southampton, United Kingdom}

\author{Derek~F.~Jackson~Kimball}
\email{derek.jacksonkimball@csueastbay.edu}
\affiliation{Department of Physics, California State University - East Bay, Hayward, California 94542-3084, USA}

\begin{abstract} A ferromagnetic gyroscope (FG) is a ferromagnet whose angular momentum is dominated by electron spin polarization and that will precess under the action of an external torque, such as that due to a magnetic field. Here we model and analyze FG dynamics and sensitivity, focusing on practical schemes for experimental realization. In the case of a freely floating FG, we model the transition from dynamics dominated by libration in relatively high externally applied magnetic fields, to those dominated by precession at relatively low applied fields.
Measurement of the libration frequency enables in situ measurement of the magnetic field and a technique to reduce the field below the threshold for which precession dominates the FG dynamics. We note that evidence of gyroscopic behavior is present even at magnetic fields much larger than the threshold field below which precession dominates. We also model the dynamics of an FG levitated above a type-I superconductor via the Meissner effect, and find that for FGs with dimensions larger than about 100 nm the observed precession frequency is reduced compared to that of a freely floating FG. This is akin to negative feedback that arises from the distortion of the field from the FG by the superconductor. Finally we assess the sensitivity of an FG levitated above a type-I superconductor to exotic spin-dependent interactions under practical experimental conditions, demonstrating the potential of FGs for tests of fundamental physics.
\end{abstract}


\maketitle

\section{Introduction}

Gyroscopes are valuable tools for metrology and navigation due to their sensitivity to rotations.
For example, the Gravity Probe B space mission contained several spinning spheres made of fused quartz and coated with a layer of niobium \cite{GravProbeB}. Changes in the direction of angular momentum and rate of rotation of these spheres were detected by a Superconducting QUantum Interference Device (SQUID). In a different  technique, ring laser interferometers (optical gyroscopes based on the Sagnac effect) have been used for continuous measurement of the Earth rotation and tilt \cite{Gebauer20}. Yet another approach observes gyroscopic motion due to precession of molecules, atoms and nuclei \cite{Zhang2019}. In the present work, we investigate how the intrinsic spin of electrons can play the role of a gyroscope. 

Atoms, molecules, and nuclei, that can possess angular momentum due to their rotational motion as well as due to intrinsic spin, can act as gyroscopes \cite{Kornack2005,Donley2010,Larsen2012,Ledbetter2012,Wu2018}. Atomic, molecular, and nuclear gyroscopes have proven to be particularly useful for precision tests of fundamental physics~\cite{Safronova2018}, including tests of Lorentz symmetry \cite{Brown2010,Smiciklas2011,Pruttivarasin2015}, searches for exotic spin-dependent interactions \cite{Vasilakis2009,Bulatowicz2013,Lee2018,Hunter2013}, dark matter experiments  \cite{Wu2019,Garcon2019,Abel2017,Roussy2020}, and measurements of electric dipole moments \cite{Graner2016,Andreev2018,Cairncross2017} and gravitational dipole moments \cite{Wineland1991,Venema1992,Kimball2017}. It has recently been proposed that a ferromagnet can act as a new type of gyroscope that may be particularly useful for precision tests of fundamental physics \cite{Kim16}. However, in order to realize the potential sensitivity of a ferromagnetic gyroscope (FG), it is essential to decouple the ferromagnet from the environment, e.g., from gravity, by requiring  either microgravity or some method of frictionless suspension. A promising platform for FG-based fundamental physics experiments involves levitating an FG above a superconducting surface by taking advantage of the Meissner effect \cite{Tao2019,Gie19,Tim19,Huillery2020,Vin20}. In the present work we model the dynamics of a freely floating FG and the dynamics of an FG levitated above a perfect type-I superconductor (SC). We find that the response of an FG to external torques is considerably modified in the case of an FG levitated above an SC: the precession frequency can be reduced by orders of magnitude as compared to that for a freely floating FG. Taking this effect into account, we analyze the sensitivity of an FG levitated above an SC to torques from exotic fields \cite{Safronova2018}. 

Under conditions where the angular momentum of a ferromagnet is dominated by the intrinsic spin of the polarized electrons, an applied torque is predicted to cause gyroscopic precession of the ferromagnet \cite{Kim16}. If such a ferromagnetic gyroscope (FG) can be sufficiently isolated from the environment, a measurement of the precession can yield sensitivity to torques far beyond that of other systems (such as atomic magnetometers \cite{Mitchell2020} and conventional gyroscopes \cite{ElSheimy2020}). The high sensitivity of an FG is a result of the rapid averaging of quantum noise \cite{Kim16,Band2018}. A key enabling technology for practical realization of an FG is a method of near frictionless suspension. One approach is to levitate a ferromagnet above an SC \cite{Tao2019}. Recently, there has been considerable interest and progress in development of sensors based on ferromagnets levitated above SCs \cite{Tao2019,Gie19,Tim19,Huillery2020,Vin20}.

Magnetization dynamics of the ferromagnet, including precession and nutation motions, have been observed in thin films using ferromagnetic  resonance \cite{Burn20,Neeraj20}. Such dynamics occur on picosecond time scales, until the intrinsic spins relax to an equilibrium state. The FG concept concerns the intrinsic spins after aforementioned relaxation, allowing to probe their dynamics, coupled to the dynamics of the ferromagnet as a whole, for times at least minutes long.

In the present work, we propose a strategy for a proof-of-principle experiment aimed at observing FG precession, and analyze a concrete example of such an experiment involving a levitating sphere above a type-I SC. We model the behavior of an FG levitated above an SC and compare to the behavior of a freely floating FG. Qualitative and quantitative differences are observed in the precession dynamics of the FG in the two cases. In relation to tests of fundamental physics, FGs have recently been proposed as tools to measure general-relativistic precession \cite{Kim20}; here we extend this discussion to show how FGs can be used in other searches for new physics.

As discussed in Ref.\,\cite{Kim16}, in order for a ferromagnet to exhibit spin precession in an applied magnetic field, it should be in the regime where the intrinsic spin $S$ due to the magnetization exceeds the classical rotational angular momentum $L$ associated with the physical rotation of the ferromagnet, $S \gg L$.
In the opposite case, where the orbital angular momentum associated with precession exceeds that of the spin along the axis, the ferromagnet ``tips over'' or, in the undamped case, oscillates or librates about its equilibrium orientation along the applied magnetic field.
These two regimes can be identified as the precessing regime and the tipping regime. Ferromagnetic compass needles operate in the tipping regime --- they tip along the direction of the external magnetic field. Atomic and nuclear spins are in the precessing regime --- they precess around the direction of the external magnetic field.

Let us reformulate the criterion $L \ll S$ for a ferromagnet to be in the precessing regime in the following way: the product of the moment of inertia $I$ and the precession frequency $\Omega$ (that represents the classical rotational angular momentum of the system) should be smaller than the spin content of the ferromagnet
\begin{align} \label{eq:precess}
 I \Omega \ll N \frac{\hbar}{2} \, ,
\end{align}
where $N$ is the number of polarized spins and $\hbar$ is Planck’s constant, such that each electron has an intrinsic spin of $\hbar / 2$.
Rephrasing \eqref{eq:precess} as a bound on frequency, we have
 \begin{align} \label{thresholdfreq}
  \Omega \ll \Omega^* =\frac{N \hbar}{2 I} \, ,
 \end{align}
 or as a bound on the external magnetic field $B$ applied on the ferromagnet,
 \begin{align} \label{BoundB}
   | B | \ll B^* =  \left| \frac{ \hbar \Omega^*}{g \mu_B} \right| \, .
 \end{align}
Here $g$ is the Landé g--factor and $\mu_B$ is the Bohr magneton. If the applied magnetic field $B$ is smaller than $B^*$, we expect the ferromagnet to be in the precessing regime. 
 
One of the key features of an FG is the fact that a torque on the electron spins generates macroscopic rotation of the ferromagnet. This behavior of an FG is closely related to the Barnett \cite{Barnett1915} and Einstein-de Haas \cite{Einstein1915,Richardson1908} effects. In contrast to nuclear spins, whose precession is largely decoupled from the crystal lattice as observed in solid-state nuclear magnetic resonance (NMR) experiments, in a ferromagnet there is strong coupling between electron spins and the crystal lattice via the exchange interaction. These internal dynamics governing an FG are well-described by the Landau-Lifshitz-Gilbert model \cite{Landau1935,Gilbert2004}.
Thus, when the electron spins within the ferromagnet are made to precess, the entire ferromagnet rotates.
 
Constructing an FG by creating suitable conditions for a magnet to precess instead of tipping opens the possibility of a sensitive measurement device. For instance, by bringing a SQUID near the FG and measuring the change in magnetic flux as the FG precesses, the torques acting on the FG can be precisely measured \cite{Kim16}. 
\begin{figure*} [tb]
\includegraphics[width=1\textwidth]
{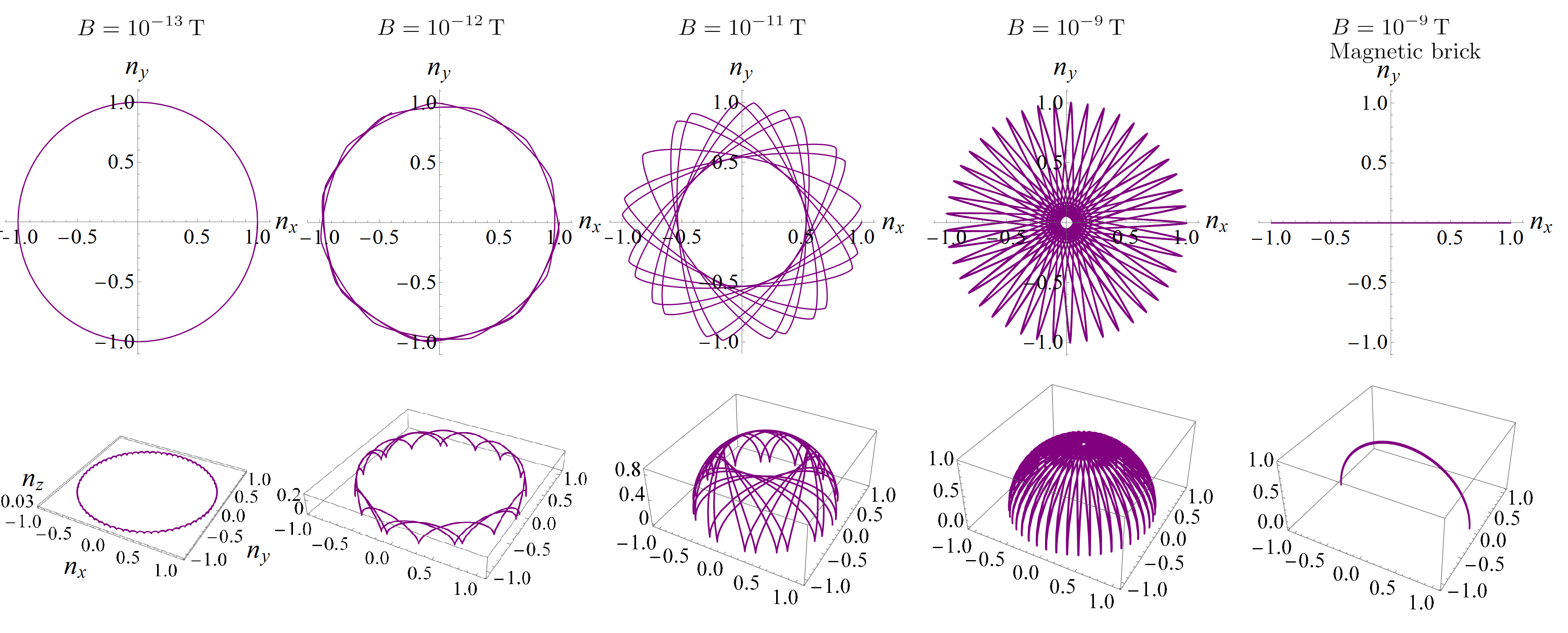}
\caption{Precession and nutation motions of a ferromagnet in a external magnetic field $B$ whose direction is perpendicular to the plane. The modelled ferromagnet has a radius of $30 \, \mu$m and contains $7 \times 10^{15}$ electron spins. Depicted is the spin vector $\v{n}$ of initial position along the $x$ axis, whose projection onto the x-y plane is shown in the upper row. The lower row shows the motion in three dimensions. As $B$ grows the precession interweaves with nutation such that the latter dominates, resulting eventually in a librational mode around the direction of $B$. For the depicted ferromagnet, the threshold magnetic field $B^*$ [Eq.\,\eqref{BoundB}] below which precession motion is dominant compared to libration, is $7 \times 10^{-12}$ T. The last column depicts the case of a ``magnetic brick", a hypothetical ferromagnet with zero spin polarization but equivalent magnetization.}
\label{nynxnz}
\end{figure*}
 
\section{Model of a freely floating ferromagnetic gyroscope} \label{freelyfloatingFG}
To better understand the dynamics of an FG, we model a freely floating FG in space subjected to a constant magnetic field $\bf{B}$, similar to the modelling in Ref.\,\cite{Kim20}.
A weak magnetic field causes precession of the FG with Larmor frequency
\begin{align} \label{omL}
    \omega_L =\frac{g_e \mu_B}{\hbar} B = \gamma B \, ,
\end{align}
where $g_e$ is the electron g--factor, and $\gamma$ is the gyromagnetic ratio.
We consider a spherical FG with radius of $30\, \mu$m and $7 \times 10^{15}$ electron spins, identical to the microsphere used in the experiment described in Ref.\,\cite{Vin20}. Note that the threshold precession frequency $\Omega^*$ described in Eq.\,\eqref{thresholdfreq} is equal to the   Einstein-de Haas frequency
\begin{align}
    \omega_I = \frac{S}{I} = \frac{N \hbar}{2 I} \, ,
\end{align}
where $I= 2 m r^2 / 5$ is the moment of inertia for a sphere of mass $m$ and radius $r$. The frequency $\omega_I$ plays the role of the nutation frequency in the zero magnetic field limit for the FG dynamics.

The equations of motion for the ferromagnet are
\begin{align} \label{dj}
    \frac{\partial \bf{j}}{\partial t} &= \omega_L \left( \bf{n} \times \bf{\hat{B}} \right) \, ,\\  \label{dn}
     \frac{\partial \bf{n}}{\partial t} &= \omega_I \left( \bf{j} \times \bf{n} \right) \, ,
\end{align}
where we defined the following dimensionless vectors: the
unit spin ${\bf n} \equiv {\bf S}/S$, the rotational angular momentum ${\boldsymbol \ell} \equiv {\boldsymbol L}/S$, and the
total angular momentum ${\bf j} = {\bf n} + {\boldsymbol \ell}$. Equations \eqref{dj} and \eqref{dn} are derived from Landau-Lifshitz-Gilbert equations under the assumption that the spin vector is locked to the easy axis, as was done in the modelling of the FG in \cite{Band2018,Kim20}. 

Solving for ${\bf{j}}(t)$ and ${\bf{n}}(t)$, Fig.\,\ref{nynxnz} shows the different kinds of motions of a freely floating FG in an external magnetic field. For magnetic fields below the threshold in Eq.\,\eqref{BoundB} the precession motion is prominent. In the intermediate regime $B \approx B^*$ both precession and nutation manifest.
At fields much larger than $B^*$ the amplitude of the nutation grows so large that it manifests as oscillation, i.e. libration, of the ferromagnet about the direction of the applied magnetic field.
Note that even in the case where libration is the dominant motion, precession of the plane of libration can still be observed.
The frequencies observed in the periodic FG dynamics in each regime can be obtained by analytical approximate solutions of the equations of motion \cite{Rusconi2017}.

 \section{Experimental strategy}
\begin{figure} [tb]
\includegraphics[width=0.5\textwidth]
{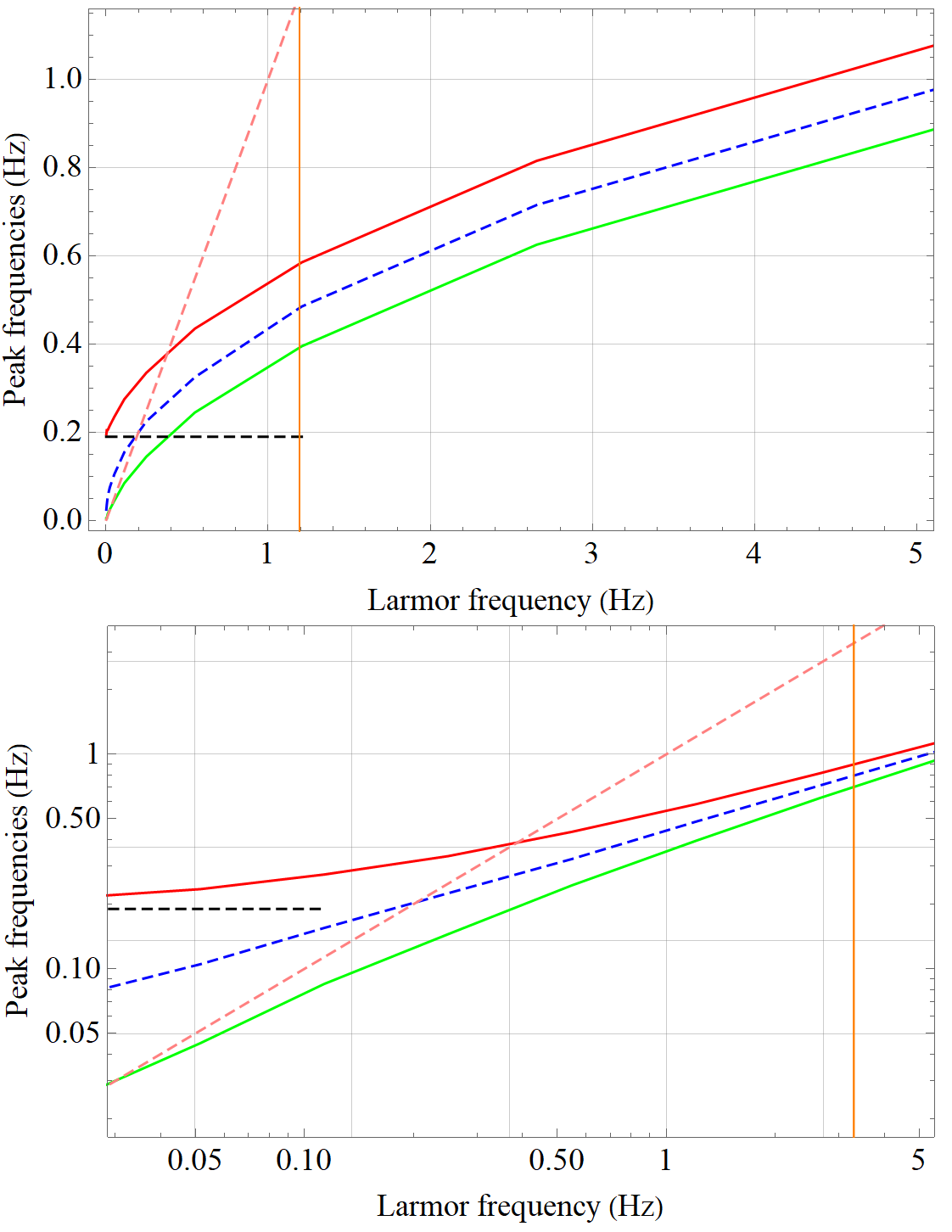}
\caption{Modelling the dynamics of an FG by Eqs.\,\eqref{dj} and \eqref{dn}. In linear and logarithmic scales, presented are the frequencies of the maxima in the spectrum of FG dynamics as can be measured with a SQUID pick-up loop, as a function of the Larmor frequency $\omega_L$.  The external magnetic field direction is perpendicular to the precession plane, as in Fig.\,\ref{nynxnz}. The SQUID pick-up loop measures the flux from the FG in the horizontal direction $x$. The middle line (dashed blue) is the sole frequency appearing in the spectrum of a ``magnetic brick" (hypothetical ferromagnet with zero spin polarization but equivalent magnetization, see main text) with a radius of $30 \, \mu$m in an external magnetic field. The red and green curves are the frequencies of a fully spin-polarized ferromagnet with the same radius, corresponding respectively to nutation and precession frequencies. 
As might be expected from Eq.\,\eqref{omegal}, the blue line is the geometric average of the red and green lines, above the threshold frequency.
The orange vertical line is the threshold frequency $\Omega^*$. 
The dashed pink line is the Larmor frequency of Eq.\,\eqref{omL}.
The dashed black line, $\omega_I$, corresponds to the nutation frequency in the zero magnetic field limit.
}
\label{splitting1}
\end{figure}

To observe precession of a ferromagnetic gyroscope, we propose to work at an external magnetic field weaker than the threshold, below which precession dominates (at sufficiently low magnetic fields, the amplitude of nutation becomes relatively small so that the dynamics of the FG are dominated by the precession). Generally, in experiments, this will require both shielding and careful control of the external magnetic field. Fortunately, the ferromagnet itself can be used as a magnetometer even for fields larger than the threshold field for precession by measuring the libration frequency $\omega_l$. Oscillation of a ferromagnet at the libration frequency $\omega_l$ was observed in soft ferromagnetic levitating particles \cite{Huillery2020} (denoted there as $\omega_\phi$) and with ferromagnets levitating above type-I SC \cite{Vin20}. For a hard ferromagnet, $\omega_l$ is the geometrical average of the Larmor frequency $\omega_L$ and the Einstein--de Haas frequency $\omega_I$ \cite{Gie19}
\begin{align} \label{omegal}
    \omega_l^2=\omega_L \omega_I \, .
\end{align}
Since the libration frequency $\omega_l$ depends on the magnetic field, it can be used to measure and reduce the magnetic field until precession dominates the dynamics.
For a freely floating FG, as one reduces the magnetic field below the threshold field defined in Eq.\,\eqref{BoundB}, the frequencies will split on the logarithmic scale  (Fig.\,\ref{splitting1}) such that they can be resolved in the magnetic flux spectrum measured by a SQUID pick-up loop. 

Quantitatively, solving Eqs.\,\eqref{dj} and \eqref{dn} for ${\bf{j}}(t)$ and ${\bf{n}}(t)$ for various $\omega_L$, the model of the dynamics of a freely floating FG shows signals at two distinct frequencies in the magnetic flux observed along the $x$-direction (perpendicular to the magnetic field applied along $z$).
These frequencies are fractionally split in low magnetic fields, that is, the difference between the frequencies, normalized by their geometric average, becomes bigger at lower frequencies, as shown in Fig.\,\ref{splitting1}, where the fractional behaviour is emphasized on the logarithmic scale. The apparent splitting of the nutation and precession curves on the logarithmic scale in Fig.\,\ref{splitting1} (the red and green curves) points to the transition from the librational behaviour above the threshold (the vertical line) into the precession and nutation motion below the threshold. 
The two frequencies can be viewed as a modulation of a central frequency (dashed blue line in Fig.\,\ref{splitting1}) which appears in the case of a ``magnetic brick'', a
hypothetical ferromagnet with zero spin polarization (${\bf j} = 0 \, {\bf n} + {\boldsymbol \ell}$) but equivalent magnetization. 
The concept of a magnetic brick is introduced to separate, in the model, effects due to magnetic torques from effects related to the gyroscopic nature of the ferromagnet.

As for the precession frequency, green curve in Fig.\,\ref{splitting1}, it deviates from the Larmor frequency in Eq.\,\eqref{omL}, as expected from the interplay between nutation and precession motions \cite{Morales2016}.
The Larmor frequency is a dashed pink line with a unit slope on the linear scale of Fig.\,\ref{splitting1}.
The parameters used in the model match those for the experimental setup discussed in the next section, which result in $\omega_I = 1.193 $ rad/s \cite{Vin20}. This $\omega_I$ is plotted as a dashed black horizontal line in Fig.\,\ref{splitting1}.

Above the threshold, the librational-mode frequency $\omega_l$ can be observed  and used to measure and control the magnetic field.
Using this technique the magnetic field can be tuned below the threshold field where the FG dynamics clearly display precession and nutation, demonstrating the gyroscopic behavior of a ferromagnet and confirming experimentally the prediction of precession. In the next section we examine this experimental strategy in the context of a ferromagnetic microsphere levitating above a type-I SC.

\begin{figure*} [tb]
\includegraphics[width=0.8\textwidth]
{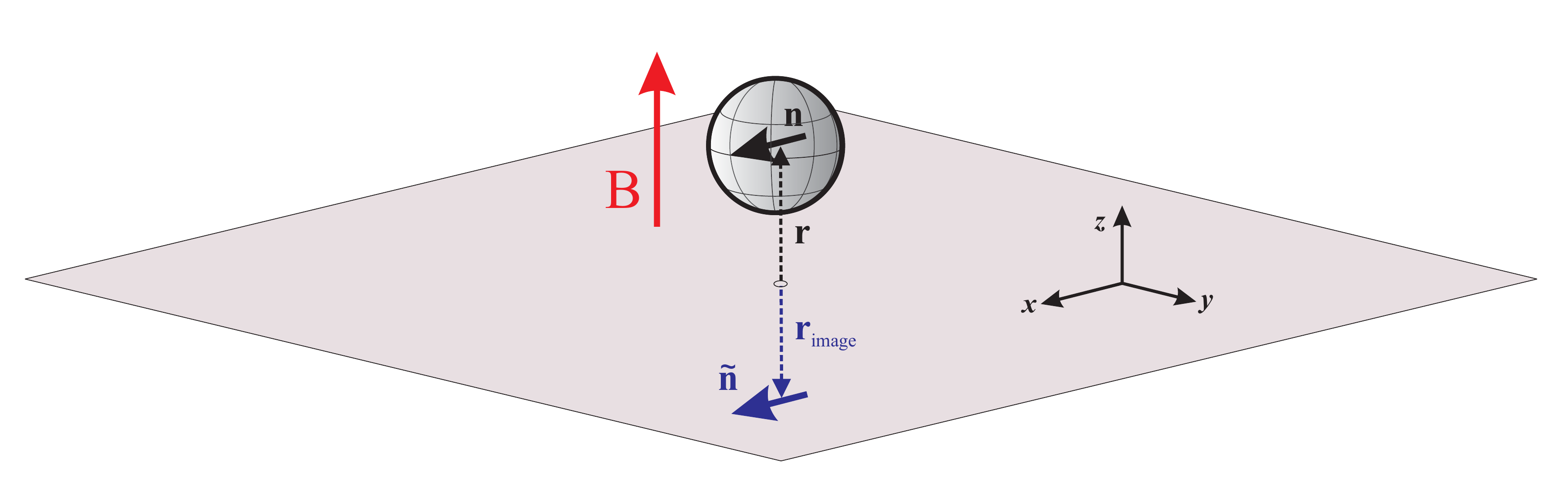}
\caption{Schematic setup for an FG levitated above a type-I superconductor. The sphere and arrow labeled $\bf{n}$ represent the FG, the gray plane represents the surface of the SC, the blue arrow labeled $\tilde{\bf{n}}$ represents the image dipole, the red arrow indicates the external magnetic field applied along the vertical direction, ${\bf r}$ is the vector pointing to the location of the center of the sphere and ${\bf r}_{\textrm{image}}$ is the vector pointing to the location of the image dipole.}
\label{FGSC}
\end{figure*}
  
\section{Ferromagnetic gyroscope levitated above a type-I superconductor}
A promising avenue for experimental realization of an FG are optomechanical and magnetomechanical systems, for instance: a ferromagnet levitated by magnetic or electric fields.
In particular, the motion, dynamics and stability of a magnetically levitated ferromagnet have been studied \cite{Rusconi2017} and are in agreement with expectations regarding the precessing and tipping regimes. Here we consider a ferromagnetic microsphere levitating above a type-I SC (Fig.\,\ref{FGSC}). 
In this case, the expulsion of the magnetic field from the superconductor by the Meissner effect creates a field in the region above the superconductor mathematically equivalent to that from image dipole. 
The image--dipole magnetic field pushes the microsphere up while gravity pulls it down. To investigate the effect of the superconductor on the FG dynamics, we include the field from the image dipole in the modelling. 

The image field $\mathfrak{B}$ is a magnetic field emanating from the image dipole located at a vertical distance $2z$ (center to center) below the levitating FG, where $z$ is the height of the FG above the SC plane. In SI units,
\begin{align}
\mathfrak{B} &= -\frac{\mu_0}{4 \pi} \frac{\mu}{\tilde{r}^5} 
\left\{
3 \, \tilde{\bf{r}} \, (\tilde{\bf{r}} \cdot \tilde{\bf{n}}) -\tilde{\bf{n}} \, \tilde{r}^2
\right\}
 \, ,
\end{align}
where $\mu_0$ is the permeability of free space, $\tilde{\bf{r}}$ is relative to the position of the image dipole
\begin{align}
& \tilde{\bf r} = {\bf r} -{\bf r}_{\textrm{image}}  \, , \nonumber  \\
& {\bf r} = \left( x , y , z \right) \, ,  \nonumber \\
& {\bf r}_{\textrm{image}} = \left( x , y , -z \right) \,  ,
\end{align}
$\mu \tilde{\bf{n}}$ is the magnetic moment of the image dipole and $\mu \bf{n}$ is the magnetic moment of the levitating ferromanget
\begin{align}
& {\bf{n}} = \left( n_x , n_y , +n_z \right) \, , \nonumber \\
& \tilde{\bf{n}} =  \left( n_x , n_y ,-n_z \right) \, .
\end{align}
Here we take the origin of the coordinate system to be on the SC plane. The image dipole has the same horizontal component of the magnetic moment as the FG, and opposite vertical component. 

To include $\mathfrak{B}$ into the equations of motion in section \ref{freelyfloatingFG}, we derive the Larmor frequency associated with this image dipole field
 \begin{align}
     \omega_\mathfrak{B} = \gamma \mathfrak{B} \, ,
 \end{align}
 so that Eq.\,\eqref{dj} contains the term
  \begin{align}
     \omega_\mathfrak{B} \left( \bf{n} \times \hat{\mathfrak{B}} \right) \, .
  \end{align}
 Moreover, we include the ferromagnet center-of-mass equations of motion
 \begin{align} \label{dptd}
    \frac{\partial \bf{p}}{\partial t} &= \frac{\mu}{2} \nabla \left(\mathfrak{B} \cdot \bf{n} \right) + m\bf{g} \, , \\ 
     \frac{\partial \bf{r}}{\partial t} &= \frac{\bf{p}}{m} \, ,
\end{align}
where $\bf{p}$ is the FG center-of-mass momentum, $\bf{g}$ is the gravitational acceleration, and the factor $1/2$ is a consequence of the image dipole being not frozen in type-I SC, i.e., following the levitating dipole \cite{Cansiz2005,Giaro1990,Lin2006}.

\begin{figure*} [tb]
\includegraphics[width=1\textwidth]
{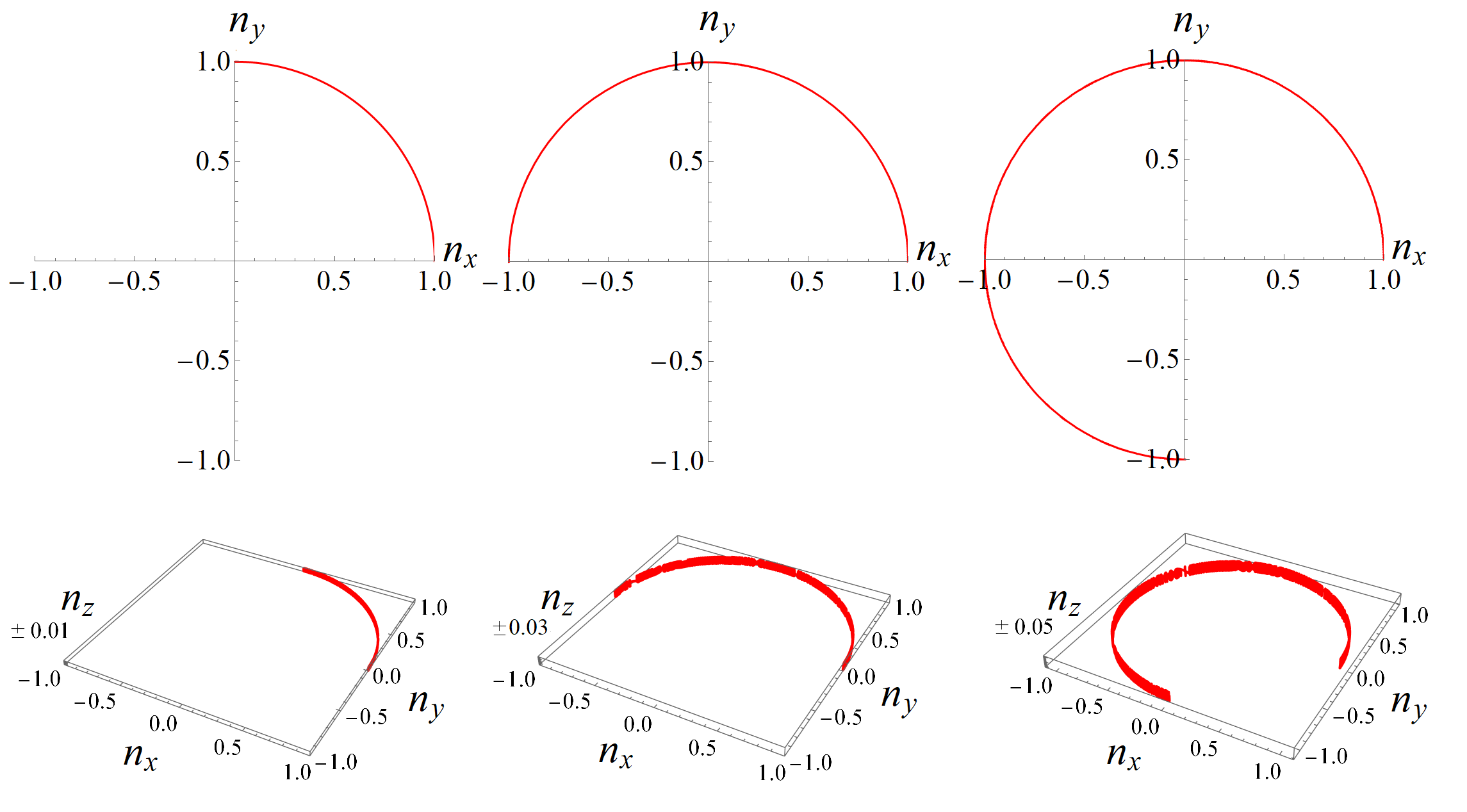}
\caption{Precession of an FG levitating above an SC. The modelled ferromagnet has a radius of $30 \,\mu$m and consists of $7 \times 10^{15}$ electron spins. Depicted is the spin vector $\v{n}$, whose projection onto the $xy$ plane is shown in the upper row. The lower row shows the motion in three dimensions. The columns, from left to right, are for tilt angles [Eq.\,\eqref{tiltangle}] of 1, 2, and 3 degrees, respectively. In the modelling time runs for $\approx 75$ seconds, which is a quarter of a period for the leftmost column, according to Eq.\,\eqref{wxy}.}
\label{imageprecession}
\end{figure*}

Modelling the levitating FG dynamics of the spin vector $\bf{n}$ and the center-of-mass motion, we recover the frequencies $\omega_z$ and $\omega_\beta$ experimentally observed in Ref.\,\cite{Vin20}, which are oscillation of the center-of-mass in the vertical direction, and libration of the magnetic moment in the vertical direction.
The dynamics of the levitated ferromagnet at the frequency $\omega_\beta$ are predominantly the result of libration caused by the image dipole field. Before introducing an additional external magnetic field to observe the effects of Larmor precession, let us note the precession motion that exists even without the introduction of an external magnetic field. We observe in the modelling a precession in the horizontal plane, with a frequency of 
\begin{align} \label{wxy}
    \omega_{xy} = \omega_I n_{z0} \, ,
\end{align}
where $n_{z0}$ is the initial vertical component of the FG magnetic moment, which is linked to the tilt angle $\beta$
\begin{align} \label{tiltangle}
    \sin{\beta} = \frac{n_{z0}}{n_0} \, ,
\end{align}
where $n_0$ is the unit spin vector and $n_z$ is the length of its vertical component, at the initial moment of the modelling.
Such a vertical component of the magnetic moment and spin translates to a vertical component of the total angular momentum, since ${\bf j} = {\bf n} + {\boldsymbol \ell}$. The librational mode $\omega_\beta$ corresponds to an oscillation between ${\bf n}_z$ and ${\boldsymbol \ell}_z$.
The image field does not change
${\bf j}_z ={\bf n}_z + {\boldsymbol \ell}_z$, thus as long as the mean value of ${\boldsymbol \ell}_z$ is not zero, precession occurs around the vertical axis, i.e. rotation of $\v{n}$ in the horizontal plane ensues. In Fig.\,\ref{imageprecession} we present examples of such a precession in the modelling. We also observe in the modelling that impacting initial angular momentum to the FG of ${\boldsymbol \ell}_z = - \sin{\beta}$ counteracts the effect of the tilt, as expected from conservation of angular momentum. 

We can explain the appearance of $\omega_{xy}$ in terms of the image dipole.
The image dipole precesses with the FG so the component of the image field in the horizontal $xy$ plane acting on the FG is constant in the rotating frame. On the other hand, the tilt of the FG with respect to the vertical axis changes the field acting on the FG due to the image dipole --- the librational oscillation causes an oscillating field along $z$ that induces FG precession. Since the librational oscillation frequency is fast compared to the precession frequency, effectively the FG is sensitive to the average field, such that bigger librational oscillation results in bigger effective field and faster precession. Note that the vertical component of the field appears due to the initial tilt angle of the FG magnetization axis out of the horizontal plane.

Such a precession was not observed in a previous experiments as the ferromagnetic microsphere was not free to rotate in the horizontal plane, because of the SC's tilt out of the horizontal plane \cite{Vin20} (or because of frozen flux \cite{Tao2019}). This tilt (or frozen flux) introduces a preferred direction for the magnetic moment of the levitating microsphere, so that it is situated in an energetic minimum; thus the ferromagnet oscillates around this direction (with frequency $\omega_{\alpha}$ \cite{Vin20}) instead of precessing in the horizontal plane.

The above precession occurs due to the image field, while to use an FG to measure external torques we seek to observe the effect due to, for example, an external magnetic field. Therefore we introduce an external magnetic field $B_{\textrm{ext}}$ (${\bf B}_{\textrm{ext}} \equiv {\bf B}$), consequently adding $\mu \nabla \left(\bf{B} \cdot \bf{n} \right)$ to the right-hand-side of Eq\,\eqref{dptd}; Eq.\,\eqref{dj} reads
\begin{align}
    \frac{\partial \bf{j}}{\partial t} &= \omega_L \left( \bf{n} \times \bf{\hat{B}} \right)
    +
    \omega_\mathfrak{B} \left( \bf{n} \times \hat{\mathfrak{B}} \right)
    \, .
\end{align}

\begin{figure} [tb]
\includegraphics[width=0.45\textwidth]
{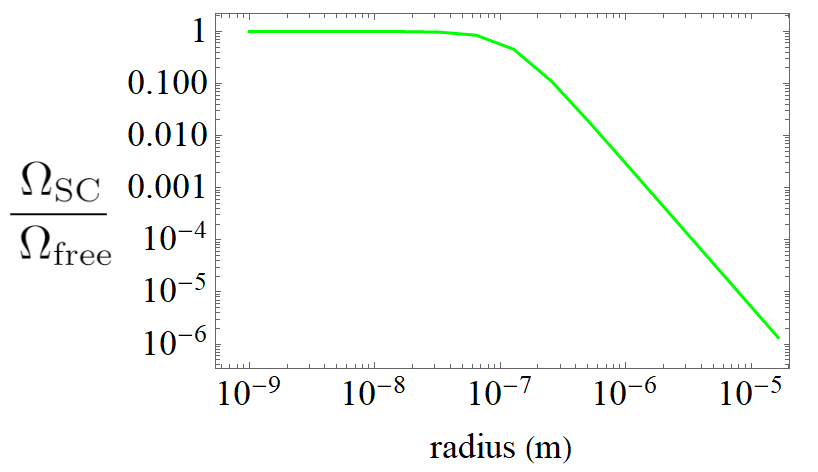}
\caption{
The ratio of a precession frequencies of an FG levitating above an SC [Eq.\,\eqref{freqSC}] to that of a freely floating FG [Eq.\,\eqref{omL}], as a function of the FG's radius. At radii below $10^{-7}$ m the ratio saturates to 1, i.e., the gyromagnetic ratio is the same as in free fall.
} \label{ratiogm}
\end{figure}

Since ${\bf j}_z$ is a constant of motion, if the tilt angle $\beta$ is initially zero (horizontal FG with respect to the SC surface), so that ${\bf j}_z(t=0)=0$, then the angular momentum associated with precession ${\boldsymbol \ell}_z = I \Omega$ must be equal and opposite to ${\bf s}_z= N \hbar \sin{\beta}$. Thus we have
\begin{align}
   \sin{\beta} = \frac{I \Omega}{N \hbar} \, .
   \label{eq:SinBetaSC}
\end{align}
The magnitude of the image dipole field a distance $z_0$ above the SC surface is
\begin{align}
    \left| \mathfrak{B} \right| = \frac{\mu \mu_0}{4 \pi z_0^3} \, ,
\end{align}
and its $z$-component is
\begin{align}
    \mathfrak{B}_z  =
   \mathfrak{B}  \sin{\beta} \, .
\end{align}
The effective magnetic field that the FG experiences is the vector sum of the external magnetic field $B_{\textrm{ext}}$ (taken to be along the $z$-axis) and $\mathfrak{B}_z$, so the precession frequency $\Omega$ is now given by
\begin{align} \label{totomega}
   \Omega &= \gamma \left(B_{\textrm{ext}}-\mathfrak{B}_z \right)
   = \gamma \left(B_{\textrm{ext}}-\mathfrak{B}  \frac{I \Omega}{N \hbar} \right) \, .
\end{align}
Solving for $\Omega$ we find
\begin{align} \label{freqSC}
   \Omega &= \frac{\gamma B_{\textrm{ext}}}{1+ \frac{\gamma \mathfrak{B}}{\omega_I}} \, .
\end{align}
Thus an FG levitated above an SC possesses an effectively reduced gyromagnetic ratio compared to the freely floating FG. The suppression of the effective gyromagnetic ratio due to the image field can be explained by a mechanism analogous to negative feedback\,\cite{Horowitz2015}.
Requiring $\Omega \ll \Omega^{*}$ for the FG to be in the precession regime gives
\begin{align} \label{SCthreshold}
  B_{\textrm{ext}} \ll \mathfrak{B} + \frac{\omega_I}{\gamma} \approx  \mathfrak{B} \, .
\end{align}
The image field is typically much larger than $B^*$; hence precession can be observed in higher magnetic fields for an FG levitated over an SC compared to a freely floating FG. However, the precession frequency is smaller compared to a free FG.  For a spherical FG with 30 micron radius, corresponding to the experiment of Ref.\,\cite{Vin20}, the ratio of a free FG precession frequency to that for an FG levitated above an SC is $\approx 4 \times 10^6$, according to the above argument. For 1 micron radius, this ratio is  $\approx 340$, so the suppression of precession frequency is reduced in the case of a smaller--radius FG. Based on Eq.\,\eqref{freqSC}, we plot in Fig.\,\ref{ratiogm} the ratio of precession rate of a levitating FG above an SC to that of a freely floating FG as a function of the FG radius.

\begin{figure} [tb]
\includegraphics[width=0.45\textwidth]
{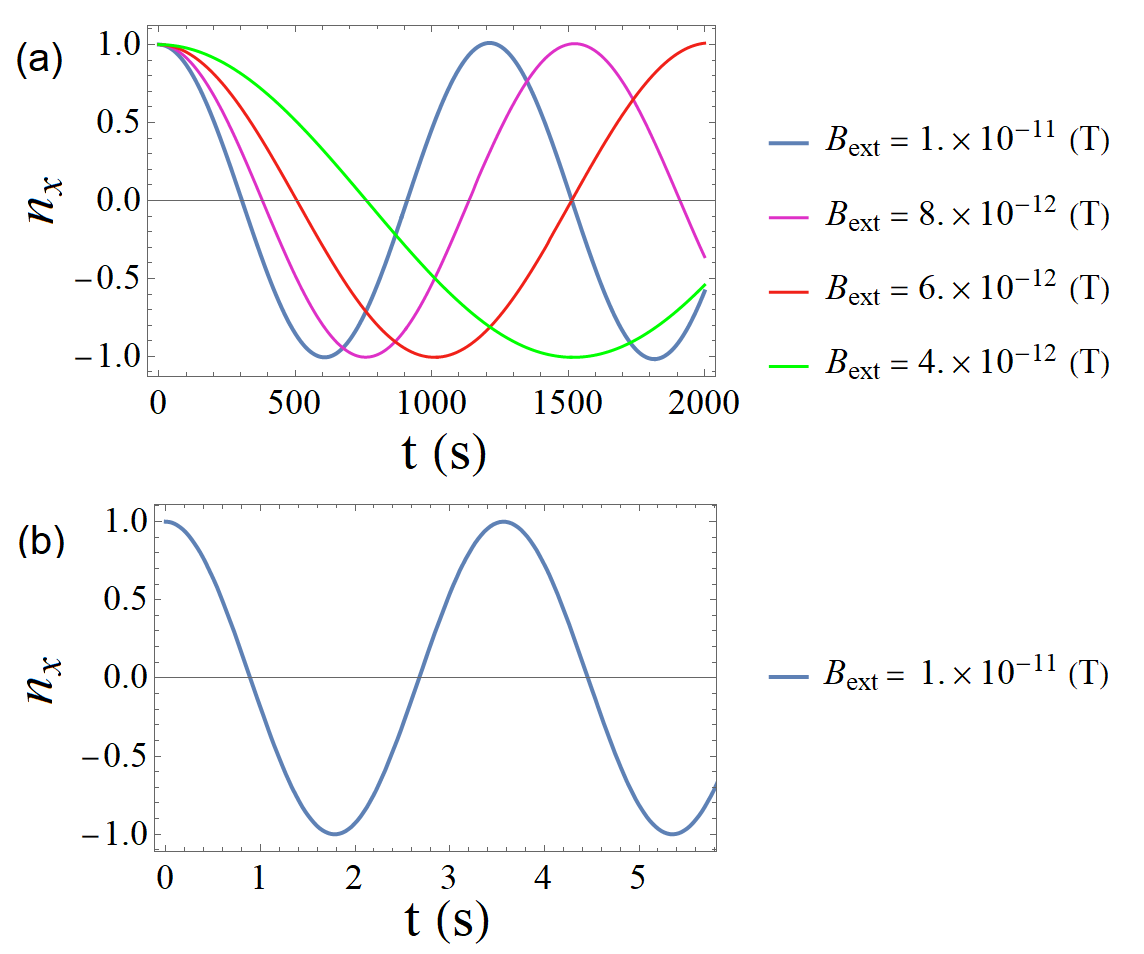}
\caption{
(a) Spin component in the horizontal direction, $n_x$, as a function of time for an FG above a type-I SC, for several external magnetic fields. The precession rates are slower from that of a freely floating FG by an amount predicted in Eq.\,\eqref{freqSC}. (b) A freely floating FG situated in an external magnetic field of $10^{-11}\,$ T. Note the time scale --- the precession frequency is about 340 times greater than for an FG above an SC for the same magnetic field in part (a).
} \label{FGtimedomain}
\end{figure}

This constraint on $B_{\textrm{ext}}$ can be viewed as an effective threshold field for a levitated FG above an SC due to the image field. 
Modelling such a system for the conditions of the levitated ferromagnet from \cite{Vin20}, but with 1 micron radius instead of 30 microns, we observe a gyroscopic behaviour in the time domain, as shown in Fig.\,\ref{FGtimedomain}, consistently $340$ times slower, for various external magnetic fields, than that in the case of precession of a freely floating FG. As another check of the negative feedback explanation, we have varied the magnitude of the image field $\mathfrak{B}$ (by varying the gravitational field magnitude) and observed in the modelling suppressed precession rates (compared to free fall FG) matching the expected rates from Eq.\,\eqref{freqSC}. In Fig.\,\ref{FGtimedomain} we decoupled the motion of the center of mass from that of the spin vector $\v{n}$, for clarity.
In the presence of both $B_{\textrm{ext}}$ and a finite initial tilt angle $\beta$ [that of  Eq.\,\eqref{tiltangle}], the resulting precession frequency is the difference between the precession frequency in the case of initial tilt angle with null $B_{\textrm{ext}}$, and that with $B_{\textrm{ext}}$ and null initial tilt angle [in accordance with Eq.\,\eqref{totomega}].

To observe $\omega_l$, and then precession, an external magnetic field should be introduced, and several modifications should be made to the  experimental apparatus used in \cite{Vin20}. One challenge is the trapping of the magnetization axis of the ferromagnet. In a previous attempt to construct an FG levitating above a type-II SC, the FG was trapped in the magnetic field lines due to the frozen image field of the FG in the superconductor \cite{Tao2019}. In Type-I SC, however, there is no frozen image dipole field. Yet, in a recent experiment with a levitating microsphere the ferromagnet was not free to rotate in the horizontal plane, i.e. trapping occurred due to a tilt of the trap \cite{Vin20}. In order to allow the microsphere to nutate and precess in the horizontal plane, a spherical trap, instead of a cylindrical one, could be employed. Following an observation of the horizontal precession due to the image field, an external magnetic field $B_{\textrm{ext}}$ can be introduced in the $z$ direction, much like was done in \cite{Tao2019}. This field is expected to cause a librational motion of the FG around it, with a frequency $\omega_l$ which can be detected with a sufficiently sensitive magnetic field sensor, such as a SQUID. Reducing the magnitude of $B_{\textrm{ext}}$ below the threshold in Eq.\,\eqref{SCthreshold}, a nutational motion will appear. Reducing further the magnetic field will reveal $\omega_L$. Note that the threshold in Eq.\,\eqref{SCthreshold} is larger relative to Eq.\,\eqref{BoundB} and thus is easier to control technically.

\section{Sensitivity to new physics}

An FG is a correlated system of $N$ electron spins that acts as a gyroscope with total spin $\sim N\hbar/2$. Spin projections transverse to the FG's magnetization axis fluctuate rapidly due to interaction with the crystalline lattice while, unless acted upon by an external torque, the expectation value of the total spin vector $\mb{S}$ remains fixed due to angular momentum conservation. This behavior enables rapid averaging of quantum uncertainty, opening the possibility in principle of measuring torques on electron spins with a sensitivity many orders of magnitude beyond the present state-of-the-art \cite{Kim16,Kim20}. For this reason, FGs can be powerful tools to search for physics beyond the Standard Model \cite{Safronova2018}.

Sensitivity estimates carried out in Refs.\,\cite{Kim16,Kim20} assume a freely floating FG in ultrahigh cryogenic vacuum at temperatures $\approx 0.1\,{\rm K}$ (residual He vapor density $\approx 10^{3}\,{\rm atoms/cm^3}$). Here we carry out sensitivity estimates for an FG levitated above a type-I SC under the vacuum conditions achieved in the experiment of Ref.\,\cite{Vin20} (residual helium pressure $\approx 10^{-5}$\,mbar, corresponding to a He vapor density of $n \approx 3 \times 10^{13}\,{\rm atoms/cm^3}$) at a temperature of $\approx 4\,{\rm K}$. We assume a spherical FG with radius $R \approx 1\,\mu$m. Therefore the conditions assumed in the following discussion are practically realizable with existing experimental apparatus with relatively minor modifications. 
In Ref.\,\cite{Vin20}, the dominant source of noise comes from collisions of He atoms with the FG. These collisions transfer angular momentum to the FG and cause a random walk of precession angle $\phi$ \cite{Kim16,Kim20}. For a spherical freely floating FG, the uncertainty in the precession frequency caused by gas collisions is given by \cite{Kim20}
\begin{align}
  \Delta \Omega\ts{col} \approx \frac{m R^2}{6 N \hbar} \sqrt{\frac{n v\ts{th}^3}{\pi t}}\,,  
  \label{eq:collisions-formula}
\end{align}
where $v\ts{th}$ is the mean thermal velocity of the residual gas atoms and $m$ is their mass. The effect of the gas collisions on FG dynamics are reduced as compared to the freely floating FG by the ``negative feedback'' from the image dipole field, under the conditions considered here by a factor of $\approx 1/340$, resulting in an uncertainty in the measured FG precession frequency of
\begin{align}
  \Delta \Omega\ts{col} \sim \frac{10^{-5}}{\sqrt{t}} \, \frac{\textrm{rad}}{\textrm{s}} \, .  
  \label{eq:collisions}
\end{align}

Other potential noise sources, such as thermal currents and blackbody radiation, were considered in Refs.\,\cite{Kim16,Kim20} and are also found to be negligible under the  experimental conditions of Ref.\,\cite{Vin20}.
Furthermore, the experimental results in Ref.\,\cite{Vin20} for a 30-micron-radius levitated ferromagnet showed that eddy current damping was negligible, and eddy current damping contributes even less for smaller FG radii: the eddy current power dissipation in a conducting sphere is $\propto r^5$. A one-micron radius ferromagnet can be single domain, in which case direct (hysteresis-based) magnetic losses should be largely suppressed as well.

The precession of the FG can be measured with a SQUID. For a SQUID with pick-up loop radius of $\approx 1~\mu$m situated about a micron from an FG (such that the flux capture is maximal), the amplitude of the time-varying magnetic flux is 
$\Phi \approx 10^{-12}\,{\rm T \cdot m^2}$.
Low-temperature SQUIDs have a flux sensitivity of $\delta \Phi \lesssim~10^{-21} \, {\rm T \cdot m^2 / \sqrt{ Hz }}$
\,\cite{Aws88,Use11,Hub08,Cla04v1}, which yields a corresponding sensitivity to the precession angle of $\delta \phi \approx \delta \Phi/\Phi \approx 10^{-9}\,{\rm rad/\sqrt{ Hz }}$. Thus the detection-limited uncertainty in a measurement of the FG precession frequency $\Omega = d\phi/dt$ integrating over a time $t$ is
\begin{align}
  \Delta \Omega\ts{det} \sim 10^{-9}\,\frac{1}{t^{3/2}} \, \frac{\textrm{rad}}{\textrm{s}}  \, .  
  \label{eq:SQUIDsensitivity}
\end{align}
Since the uncertainty in the measurement of precession due to gas collisions is far larger than that related to the SQUID measurement, requirements on the pick-up loop geometry and SQUID sensitivity are correspondingly relaxed.

Estimates have shown that $\Delta \Omega\ts{det}$ appears to surpass the ``standard quantum noise limit" \cite{Kim16,Giovannetti04}. While the energy resolution per bandwidth ($E_R$) for existing magnetometers is at or above the quantum limit $\hbar$, an FG can in principle achieve $E_R \ll \hbar$, under conditions where external sources of error are controlled so that the FG sensitivity is limited by detector noise \cite{Mitchell2020}. Such accuracy arises as the quantum uncertainty is rapidly averaged by the internal ferromagnetic spin-lattice interaction, while the FG maintains gyroscopic stability due to the conservation of the total angular momentum, which is dominated by the intrinsic spin. Another way to understand the sensitivity of an FG is to note that the ferromagnetic spin-lattice interaction spreads the quantum fluctuations over a broad frequency band ($\gtrsim 1 - 100$\,GHz). Due to the gyroscopic stability, one can still measure slow changes of the overall average direction of the FG spin. Integrating over long periods of time averages the quantum fluctuations, acting as a ``low-pass filter'' for the quantum noise, and thus a high sensitivity to comparatively low-frequency spin precession can be achieved.

As an example of the potential of FGs as tools for testing fundamental physics, we consider an experimental search for yet-to-be-discovered (exotic) spin-dependent interactions mediated by new bosons \cite{Moo84,Dob06,Fad19}. In particular, axions and axionlike particles (ALPs) mediate a pseudoscalar ($P$) interaction between electrons described by the potential  
\begin{widetext}
\begin{align}
\sV_{PP}(\mathbf{r}) = \frac{(g_P^e)^2}{4 \pi \hbar c} \frac{\hbar^3}{4 m_e^2 c} \sbrk{ \mathbf{S}_1 \cdot \mathbf{S}_2 \prn{ \frac{mc}{\hbar r^2} + \frac{1}{r^3} + \frac{4\pi}{3} \delta^3(\mb{r})}  - \prn{ \mathbf{S}_1 \cdot \hat{ \mathbf{r} } } \prn{ \mathbf{S}_2 \cdot \hat{ \mathbf{r} } }  \prn{  \frac{m^2c^2}{\hbar^2 r} + \frac{3mc}{\hbar r^2} + \frac{3}{r^3} } } e^{ - mc r / \hbar }\,,
\label{Eq:pseudoscalar-interaction}
\end{align}
\end{widetext}
where $(g_P^e)^2/\prn{ 4 \pi \hbar c }$ is the dimensionless pseudoscalar coupling constant between electrons, $m_e$ is the electron mass, $\mathbf{S}_{1,2}$ are the electron spins, $m$ is the mass of the hypothetical pseudoscalar boson, $c$ is the speed of light, and $\mathbf{r} = r \hat{ \mathbf{r} }$ is the separation between the electrons. 

One could search for spin precession induced by the pseudoscalar-mediated dipole-dipole interaction,  Eq.\,\eqref{Eq:pseudoscalar-interaction}, by modulating the distance between a polarized spin source and a levitated FG. Some of the most stringent laboratory constraints on such exotic dipole-dipole interactions have been achieved using spin-polarized torsion balances \cite{Hec08,Ter15} with SmCo$_5$ as a polarized spin source. In SmCo$_5$, the orbital magnetic moment of the Sm$^{3+}$ electrons nearly cancels their spin moment, and so SmCo$_5$ possesses a high spin polarization while having a relatively small magnetic moment, thus reducing magnetic-field-related effects. The spin--polarized source in such an experiment could be positioned underneath the SC to further shield the FG from the magnetic field due to the spin source. Although the SC will shield the FG from the magnetic field of the SmCo$_5$ spin source, it turns out that the pseudoscalar interaction \eqref{Eq:pseudoscalar-interaction} is unshielded by the SC \cite{Kim16shielding}. This is a consequence of the fact that SC shielding relies on the coupling of magnetic fields to currents rather than to electron spins. Thus, since the Meissner effect is unrelated to interactions with the electron spins, the SC shield has no effect on the pseudoscalar-mediated dipole-dipole interaction described by Eq.\,\eqref{Eq:pseudoscalar-interaction} \cite{Kim16shielding,Hec08}. 

\begin{figure} [tb]
\includegraphics[width=0.48\textwidth]
{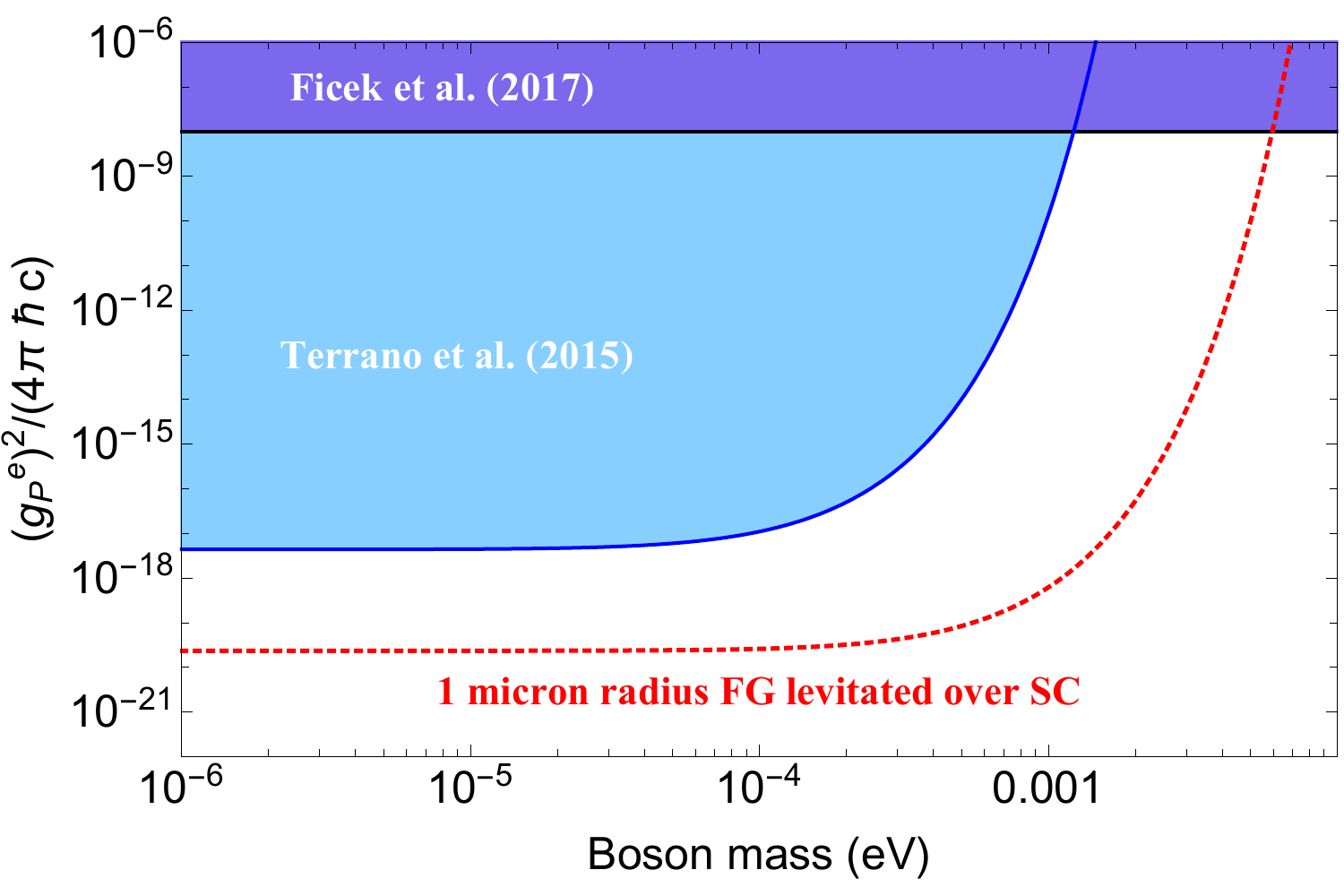}
\caption{
Comparison between the existing experimental constraints (solid lines and shaded regions) on a pseudoscalar-mediated dipole-dipole interaction between electron spins and the projected sensitivity of an experiment using a one micron radius spherical FG levitated above a type-I SC (dotted red line). Constraints shown with the black line and dark blue shaded region are based on He spectroscopy \cite{Fic17}; constraints shown with the blue line and light blue shaded region are from an experiment using a spin-polarized torsion pendulum \cite{Ter15}. The proposed experiment with the levitated FG assumes as a polarized spin source a 1-mm radius SmCo$_5$ sphere positioned 1\,mm away from the FG and an integration time of $t = 10^6$\,s.   
} \label{DipoleDipoleConstraints}
\end{figure}

An experiment using a one--micron--radius FG levitated above an SC would be sensitive to the region of parameter space bound from below by the dotted red line in Fig.\,\ref{DipoleDipoleConstraints}.
We assume that the SmCo$_5$ spin source is a one-mm radius sphere positioned one mm away to the FG to allow space for the SC.
A one-mm-radius SmCo$_5$ sphere would contain $\sim 5 \times 10^{19}$ polarized electron spins. The FG sensitivity to spin precession is given by Eq.\,\eqref{eq:collisions}. For comparison, Fig.\,\ref{DipoleDipoleConstraints} shows the most stringent laboratory constraints in this region of parameter space, which are based on spin-polarized torsion-balance measurements \cite{Ter15} and He spectroscopy \cite{Fic17,Del17}; related experiments are discussed in the review \cite{Safronova2018} and Refs.\,\cite{Hun13,Hec13,Kot15,Sav18,Ji18,Ron18,Jia20,Din20}. Compared to these existing constraints, our proposed experiment with a levitated FG can explore many decades of unconstrained parameter space. This illustrates the potential of FGs as tools to search for exotic spin-dependent interactions, which could open a window to beyond-the-Standard-Model physics.  

\section{Conclusion}

In summary, we present a roadmap for experimental realization of a ferromagnetic gyroscope (FG). In essence, an FG is a ferromagnet that precesses under the influence of external torques. Ferromagnetic gyroscopes can be used as a new types of magnetometers and gyroscopes, and we have shown that they can be particularly useful as tools for precision tests of fundamental physics. 

We model and explain the dynamics of a freely floating FG in space. We propose a strategy to experimentally  realize an FG. The librational mode in the magnetization dynamics serves as a calibration tool for the applied magnetic field. This enables the magnetic field to be tuned such that the precession mode of an FG is observable. 

We also compare the dynamics of a freely floating FG to that of an FG levitated above a type-I superconductor. The effect of the SC is modelled using an image dipole field. We find that the SC has a significant effect on the FG dynamics: the image dipole field generates a ``negative feedback'' that effectively suppresses the gyromagnetic ratio as compared to the case of a freely floating FG. The effective magnetic field threshold below which precession motion is dominant is higher for the case of an SC compared to a freely floating FG [Eq.\,\eqref{SCthreshold}] while the observed precession frequency for a given field strength is lower [Eq.\,\eqref{freqSC}].

~\

 \section{Acknowledgements}
P. F. would like to thank Martti Raidal and the team in NICPB for their hospitality in Tallinn, Estonia. 
This research was supported by the Heising-Simons and Simons Foundations, the U.S. National Science Foundation under Grant No. PHY-1707875, the DFG through the DIP program (FO703/2-1), and by a Fundamental Physics Innovation Award from the Gordon and Betty Moore Foundation. The work of D. B. supported in part by the DFG Project ID 390831469: EXC 2118 (PRISMA+ Cluster of Excellence), the European Research Council (ERC) under the European Union Horizon 2020 Research and Innovation Program (grant agreement No. 695405) and the DFG Reinhart Koselleck Project. The work of A. S.  supported in part by the US National Science Foundation grant 1806557, US Department of Energy grant DE-SC0019450, the Heising-Simons Foundation grant 2015-039, the Simons Foundation grant 641332, and the Alfred P. Sloan foundation grant FG-2016-6728. The work of C. T., A. V., and H. U. supported in part by the EU H2020 FET project TEQ (Grant No. 766900), the Leverhulme Trust (RPG2016-046) and the COST Action QTSpace (CA15220). 

\section{Supplementary Material}
\subsection*{Alternative Abstract}
Might a magnet be a spinning top?

Instead of pointing to one's north,

wobble around it and nod

by intrinsic spins cohort.

Aye, at small enough magnetic field

a spectrum splitting is revealed.
\\ \\

A levitating sphere above cool plate

modelled to show it in such state.

Once a compass made to be

a gyroscope with precision spree,

for exotic fields it may test

and basic laws of physics attest.

\end{document}